# Multi-platform Process Flow Models and Algorithms for Extraction and Documentation of Digital Forensic Evidence from Mobile Devices

Gilbert Gilibrays Ocen[1], Ocident Bongomin[2], Gilbert Barasa Mugeni[3], Mutua Stephen Makau[4], Twaibu Semwogerere[1]

[1]Department of Computer Engineering, Faculty of Engineering, Busitema University, Tororo, Uganda
[2]Department of Manufacturing, Textile and Industrial Engineering, School of Engineering, Moi university, Eldoret, Kenya
[3]Directorate of Innovation, Research and Development, Communication Authority of Kenya, Nairobi, Kenya
[4]Department of Computer Science, School of Computing and Informatics, Meru University of Science and Technology, Meru, Kenya

Correspondence to be assigned to Gilibert Gilibrays Ocen, gilbertocen@gmail.com

## Abstract

*The increasing need for the examination of evidence from mobile and portable gadgets increases the essential need to establish dependable measures for the investigation of these gadgets. Many differences exist while detailing the requirement for the examination of each gadget, to help detectives and examiners in guaranteeing that of any kind piece of evidence extracted/ collected from any mobile devices is well documented and the outcomes can be repeatable, a reliable and well-documented investigation process must be implemented if the results of the examination are to be repeatable and defensible in courts of law. In this paper we developed a generic process flow model for the extraction of digital evidence in mobile devices running on android, Windows, iOs and Blackberry operating system. The research adopted survey approach and extensive literature review a s means to collect data. The models developed were validate through expert opinion. Results of this work can guide solution developers in ensuring standardization of evidence extraction tools for mobile devices.*

## 1. Introduction

Attempts to extract information from several gadgets using a variety of mobile forensic tools and process models have revealed contradictory results [1–3]. Correspondingly, with computerization and improvement in usability, an extra attention should be taken in ensuring that the methods are outstandingly rigorous, forensically dependable, and certifiably correct, vigilant documentation of the steps and abilities of automated tools is essential [4]. Documented approaches permits an examiner to recall exactly the actions taken to gather information and can be used to refute claims of mishandling [5].

Scholarly works of most researchers affirms that forensic science suffers from a lack of documentation and transparency [6], Standard and well laid approaches are critical in guiding the handling of electronic evidence by examining authorities. Hence**,** the aim of the documentation is to establish that the extraction process was done in a forensically acceptable manner and given the same conditions and by following the same method will yield the same results every time [7]. Proper documentation is evidenced by records, photos, and tool-generated report content [8]. Though the investigator might thrive in extracting the expected information using available tools, additional detailing of information could be beneficial, exclusively for court proceedings [9].

Digital Forensics is a broad term that covers the retrieval and examination of material found in digital devices, often concerning computer crime [10, 11]. The role of forensic science can be perceived as using methodical approaches, actions, and investigation frameworks to extract, preserve, gather, analyze and provide scientific and technical shreds of evidence for the criminal or civil courts of law [12], and to organize good documentations for law enforcement prosecution [10, 11]. On the other hand, Digital forensic is the practice of finding, preserving, scrutinizing, and presenting evidence in a manner that is legally acceptable [12]. These definitions are supported by [13] who contends that digital evidence is viewed as material and records of importance to the examination which is kept on, delivered, or conveyed via an electronic device.

The constant industrial growths and rising popularity of mobile digital devices increase substantial challenges, circumstances and scenarios for investigators and prosecutors globally, the existence of several tools and systems with different process models make it hard for even a trained investigator to select an appropriate forensic tool for seizing internal files from mobile devices [14]. Many forensic models emphasize the examination a particular operating system platforms [15], ignoring a more critical aspect of consistency and documentation on the approaches and steps followed. Whereas [16] listed many forensic evidence preservation techniques from the view point of efficiency in the general forensic context for mining and documentation of evidence from mobile devices, little or no attempts have been done in regards to methodological documentation and the consistency of process models followed during the extraction of such pieces of information. While [17] notes that despite of the growing awareness and research on practical forensics, the explanation and implementation still lack consistency in the digital forensics community, an argument supported by recent researches such as [9, 18, 19].

The ever changing technological and industrial developments coupled with the variety of complexities brought about by today's demand for information from mobile devices make forensics investigators face serious adaptation problems, in standardizing and adopting to acceptable models that can be used to counter this rising need [20, 21]. Trustworthiness of evidence is entrenched directly on adopted examination processes, selecting to avoid or accidentally avoiding a step can lead to insufficient evidence and increase the risk of denying such in legal proceeding [22]. Currently, no standard or universally accepted process model has been developed that can be used for obtaining evidence from mobile devices and the vivid expansion of smart devices proposes that any forensic investigator will have to use as many independent models as necessary to gather and preserve information [23].

The existing models cannot address the growing demands of digital evidence arising from the increased use of mobile devices and the complexities that perpetual crime committers present while using these devices therefore since a number of these models either concentrate on a particular stage of the extraction process [24, 25] or are operating system platform dependent. Basing on existing research in digital forensic process models that can be used to acquire evidence in mobile devices. Generally, literature indicate the requirements that have been found to guide and measure the extraction process of digital evidence in mobile devices and their performances, reliability, performance, and validity are; policy, extraction method, nature of data, data type, technical documentation and forensic extraction tools.

## 2. Methodology

The present study was conducted in Four steps or phases as illustrated in Figure 1. Literature on specific email security techniques were reviewed in phase one, in phase two the algorithm was developed, and in phase three the algorithm was evaluated using questionnaires with selected participants and SWOT analysis was done in the last phase.

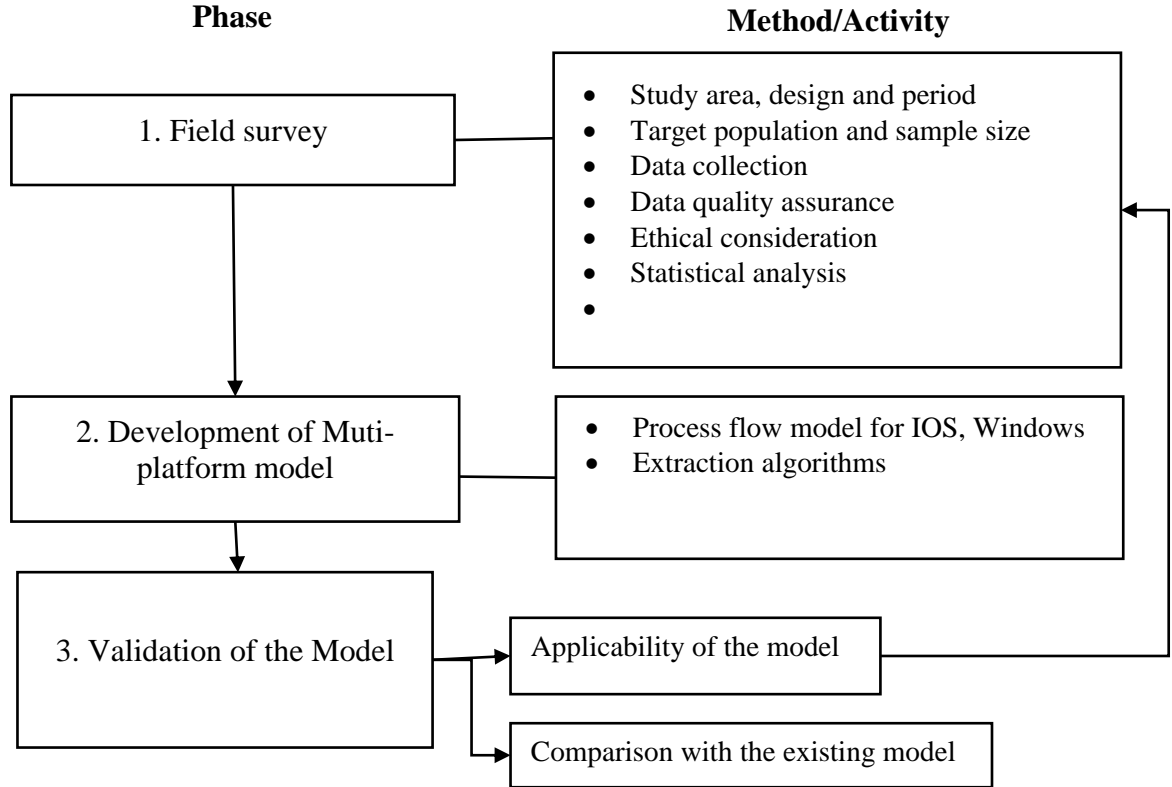

Figure 1: Methodology Approach.

### 2.1 Field survey
#### 2.1.1 Study area, design and period

The study was conducted in Kampala, Uganda because it is where the researcher found deemed to get most of the respondents with knowledge in the subject matter. From this location, the researcher was able to identify law enforcement agencies such as the police, judicial officials, computer forensic experts and practitioners, researchers in the field of computer forensic investigations and evidence extraction, mobile telecommunication, and banking sectors that have several crime/ fraud departments for investigating crimes related to use of technology. The cross-sectional survey design was used in this study for the period of one year from 2018 to 2019.

#### 2.1.2 Population and sample sizes

The study population comprised of respondents from Law enforcement agencies especially the Uganda Police Force especially the Crime Intelligence and Investigation Department (CIID), Directorate of Public Prosecution, officials of courts of law especially the Advocates, court registrars, magistrates and judges, Policy Makers, people from regulatory bodies such as; Uganda

Communications Commissions, National Information Technology Authority Uganda (NITA-U), a Business community comprising of Telecommunications operators like Mobile Telecommunication Network (MTN-Uganda), Airtel Uganda since these are the largest telecommunication service providers offering financial services, Banks such as Stanbic Bank, Centenary Bank, Barclay's bank Uganda and Standard Chartered Bank because these are the biggest providers of online transaction systems that use some of the mobile digital devices in their operations. Purposive/ judgment sampling was used alongside snowball sampling tool to compliment purposive especially in studying different operating system platforms, inconsistencies, and technical documentation of extraction process models while simple random and stratified sampling was used for probability sampling this is because the researcher collected data from different sectors and organized them into different strata and simple random sampling applied. The population sample was determined using Krejcie and Morgan [26] sampling table derived using the formula. The Krejcie and Morgan's sample size calculation was based on p = 0.05 where the probability of committing a type I error is less than 5 % or p< 0.05 [26].

**Table 1.** Sample size determination using Krejice and Morgan sampling technique

| Sector | Population size | Sample size |
| --- | --- | --- |
| Law Enforcement Agencies | 10 | 7 |
| Regulatory Authorities | 20 | 11 |
| ICT experts | 100 | 63 |
| ICT Researchers | 20 | 11 |
| Policymakers | 30 | 16 |
| Business communities | 70 | 31 |
| **Total** | **200** | **130** |

From Table 1, it is clear that the population size that the researcher considered for law enforcement agencies was ten and the sample size that the researcher used was seven, while for the majority of the respondents were drawn from information and communication technology experts (ICT) where the sample population was one hundred (100) and the researcher took the sample size of sixty-three, this was followed by the business community who comprised of people in the banking industry, telecommunication agencies where the population size was seventy and the sample size was 31. Overall, the target population size was two hundred and the sample size was one hundred and thirty.

### 2.1.3 Data Collection

In this study, two data collection tools were adopted for survey: Questionnaire and interviews. Questionnaires covered a wide range of area of the chosen populace, offered a consistent form of responses, reduced bias, did not make people apprehensive and were completed at the respondent's convenience [27]. Questionnaires were developed for the various categories of respondents like policymakers, law enforcement agencies, researchers, Information and Communication Technology (ICT) experts, regulatory authorities and business community with a view of obtaining different kinds of data from these categories of respondents. Questionnaires were developed based on the understanding derived from the literature reviewed in the areas such as mobile devices operating systems platforms, technical documentation, process model inconsistencies and complexities as independent variables, and a multi-platform digital extraction process model for mobile devices forensic evidence. The Likerts' standard scale based on five points was used to design the questionnaires in the scale of strongly agree to strongly disagree (5-1). Interviews were used to supplement the questionnaires and were tightly structured, administered mainly to Information and Communication Technology (ICT) experts within the Law enforcement agencies, policymakers, regulatory agencies, and business community as well as those in data recovery and forensic departments from agencies such as telecommunication networks, banking industry and researchers in the field of digital forensic using note-taking.

### 2.1.4 Data quality assurance

The term "reliability" is used to describe the "repeatability" or "consistency" of the measure [28], In this research, the internal consistency reliability methodology was adopted. According to Chen [29], the internal consistency method uses "a single measurement instrument administered to a group of people on one occasion to estimate reliability. The reliability of the instrument is judged by estimating how well the items that reflect the same construct yield similar results". Cronbach's Alpha coefficient (α) was chosen as the best approach to estimate the reliability of constructs by examining the internal consistency of the measure. As stated by Spencer [30], there are four kinds for reliability coefficient α; excellent reliability (α >=0.90), high reliability (0.70 < α < 0.90), moderate reliability (0.50< α< 0.70), and low reliability (α<= 0.50). All the constructs used in this study passed the reliability test as reflected in Table 2.

Table 1. Reliability Test of constructs using Cronbach's coefficient (alpha)

| Construct | No. of Items | Cronbach's Alpha |
|---|---|---|
| Policy Factors (PF) | 7 | 0.591 |
| Operating system platform (MDF) | 4 | 0.741 |
| Device factors (DF) | 4 | 0.640 |
| Extraction Method factors (EM) | 15 | 0.781 |
| Data type factors (DT) | 11 | 0.807 |
| Nature of Data factors (ND) | 5 | 0.778 |
| Forensics Extraction tools (FET) | 9 | 0.850 |
| Forensics Documentation process (FDP) | 10 | 0.640 |

In this study, Cronbach's α was 0.807 for the Forensics Extraction Tools (FET) constructs and 0.591 for Policy Factors (PF) constructs. Forensics Extraction Tools (FET) constructs had the highest reliability (α =0.850), Data Type constructs (α =0.807), Extraction Method factors (EM) (α =0.781), Nature of Data factors (ND) (α =0.778), Operating system platform (MDF) (α =0.741), While Forensics Documentation Process (FDP) and Device factors (DF) both had (α =0.640), and finally Policy Factors (PF) (α =0.591). As reported by Perry et al. [28], these figures indicate that out of the eight (8) constructs, five had high reliability while had moderate reliability. Implying that constructs were internally consistent. Therefore, all items of each construct measured the same. While validity of the instruments was determined using the Content Validity Index (CVI). It was performed on the constructs to ensure that the scale items are meaningful to the sample and capture the issues that were measured. The measuring instruments were further tested to ensure quality and validity; this was done after doing a pilot study using 30 questionnaires. The content validity indices from the three experts are 0.982, 0.964, and 0.967. Therefore, it was observed that the Content Validity coefficients were > 0.6, and hence the scales used to measure the study variables were consistent. Moreover, it is valid because for Cronbach alpha coefficient greater than 0.5 is considered moderate validity and above .90 considered excellent validity, in this research all variables posted above 0.50 which indicates good to excellent validity meaning all the constructs and the sub-indices in this study passed the validity tests.

### 2.1.5 Ethical consideration

The ethical clearance for the survey was obtained from the Institutional Research Ethical Committee of Busitema University and informed consent of respondents before enrolling them voluntarily in the study. Ethics issues such as privacy and confidentiality of the respondents were ensured [31]. Besides, the letter was acquired from the university that acted as an introductory document to different organization and individuals engaged on this research. It was also ensured that the extraction model developed does not execute any unintended/ undisclosed activity in the users' devices.

### 2.1.6 Statistical analysis

The analysis was performed using Statistical Package Software for Social Scientist (SPSS) version 20.0 (SPSS, Chicago, Illinois) and the descriptive statistics was used for extracting results from the analysis on all the study variables. Descriptive statistics were performed for the constructs used in this study to determine their significance using the mean responses to obtain their ranking as per the number of responses from the participants who conquer with literature as to whether these constructs were used in this study were contributors to inconsistencies in mobile devices evidence extraction process models. The regression analysis was performed with consistency metric (CM) as the dependent variable and constructs including Extraction methods (EM), Forensic Extraction Tools (FET), Policy Factors (PF), Device Factors (DF), Nature of Data (ND) and Data type factors (DTF)) as independent variables.

## 2.2 Model development

### 2.2.1 Multi-platform flow model

The model design and validation involved the use of business process model development, analytical hierarchy approach, experimental and experts' opinion was used to validate the developed model. An experimental setup was conducted to test the process model developed to check for consistencies in the extraction process models. The process flow for multi-platform model is depicted in Figure 2. The individuals flow model for the iOS and window mobile devices are presented in Figure 3 and Figure 4, respectively.

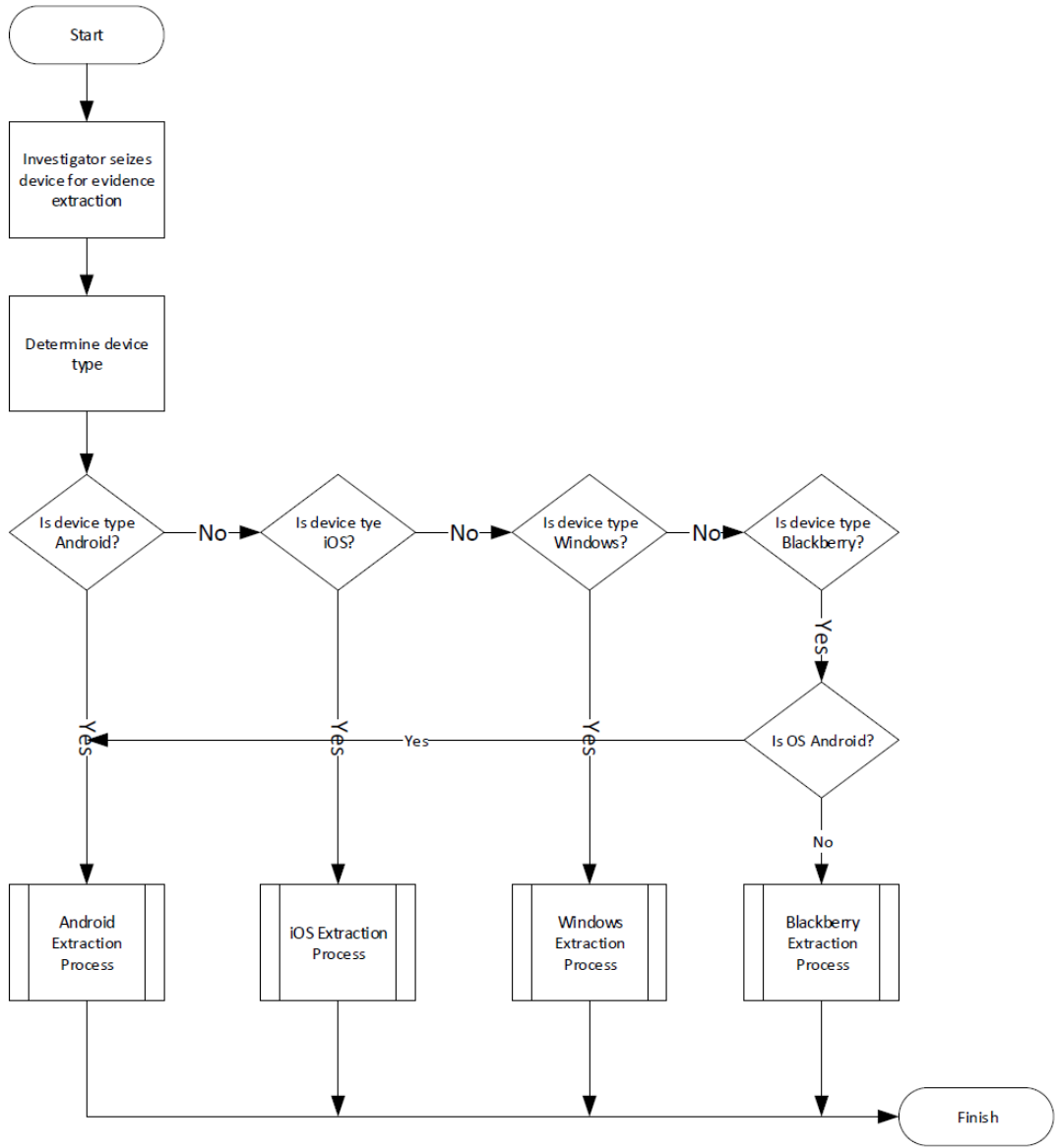

Figure 2. Process flow for multi-platform model.

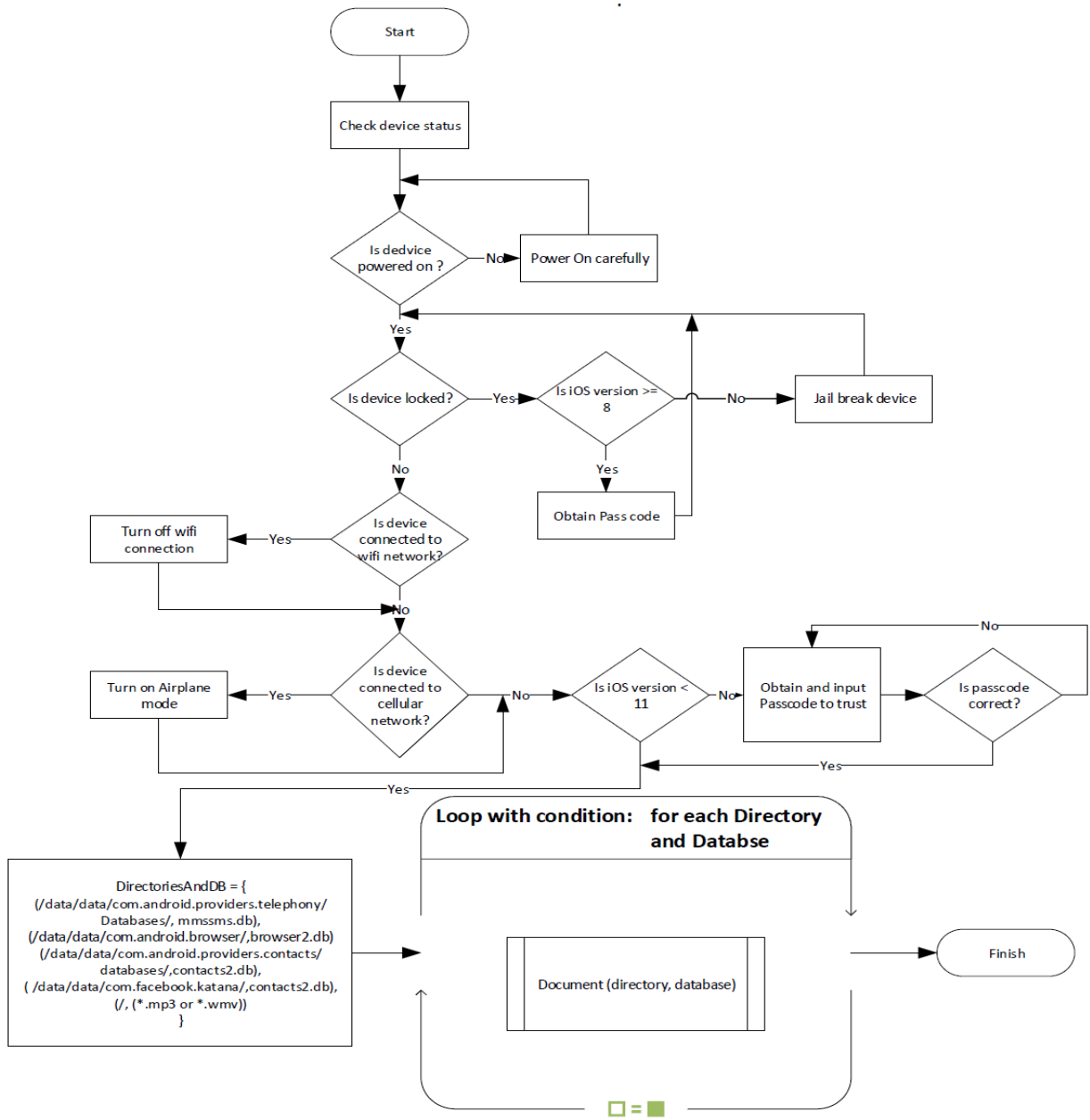

Figure 3. Process flow model for the case of IOS

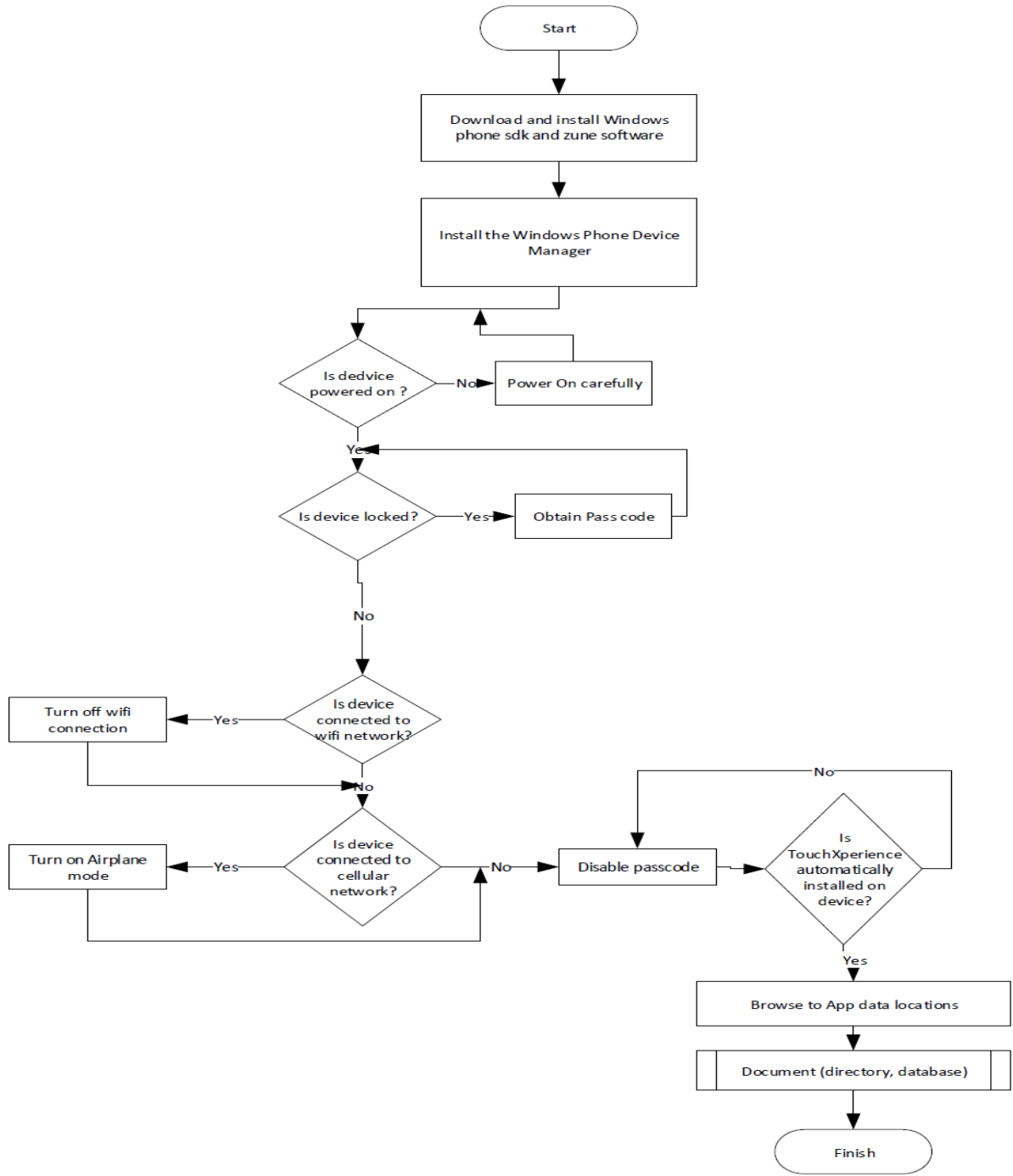

Figure 4. Process flow of Multiplatform model for the case of Window OS

### 2.2.2 Description of extraction algorithms

First and fore most the gadget is seized for evidence extraction. A check is done to ascertain the type of operating system running on it, if it's android, the android extraction process is carried out which under the ExtractFromAndroid(SiezedDevice). At the beginning, the gadget status is checked, for example; power status, Wi-Fi connection and cellular network status. This action is performed on all gadgets to ensure the gadget is powered and not having any network connection. After this check, Universal Serial Bus debugging is activated via developer options, screen timeout is increased and root access is gained. After which browsing to various directories/ locations, and getting the SQLite databases which can be opened to acquire evidence, which is later documented using Document (directory, dictionary). The procedure is followed similar steps while ensuring documentation so as to allow consistence.

In case where it is an iOS as depicted in Figure 5, ExtractFromiOS(SiezedDevice) is trailed with the same actions of having the gadget status checked, However the difference with this extraction happens when connecting to a personal computer where a trust code is required between the device and computer for cases of iOS11 and above. Documentation takes place through (directory, dictionary). During extraction from Windows devices as shown in Figure 6, ExtractFromWindows(SizdDevice) is activated that necessitates installing windows phone sdk and zune software, the windows phone device manager. The gadget status is checking is done. If there is connection of the gadget to the work station done, then automatic installation of TouchXperience on the phone is followed. This enables browsing to different directories and access the different files and the documentation is followed using Document (directory, dictionary).

Finally, for BlackBerry based gadgets there is relatively small variations with other devices, ExtractFromBlackBerry(SiezeDevice) is done, information /data acquired from backup files as opposed to the devices itself due to the fact that its security complexity. BlackBerry Desktop Software is installed and opened which detects a blackberry device and creates backup files. The files are browsed for evidence which is documented using Document (directory, dictionary).

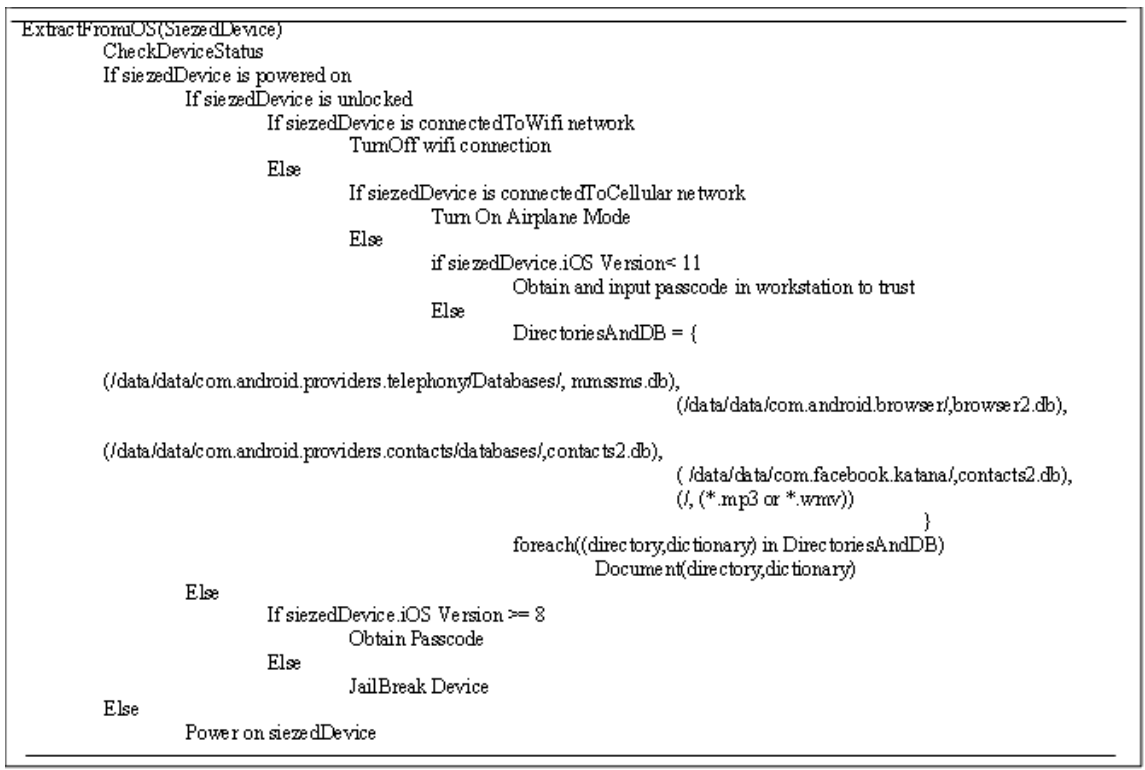

Figure 5. Extraction algorithms for IOS

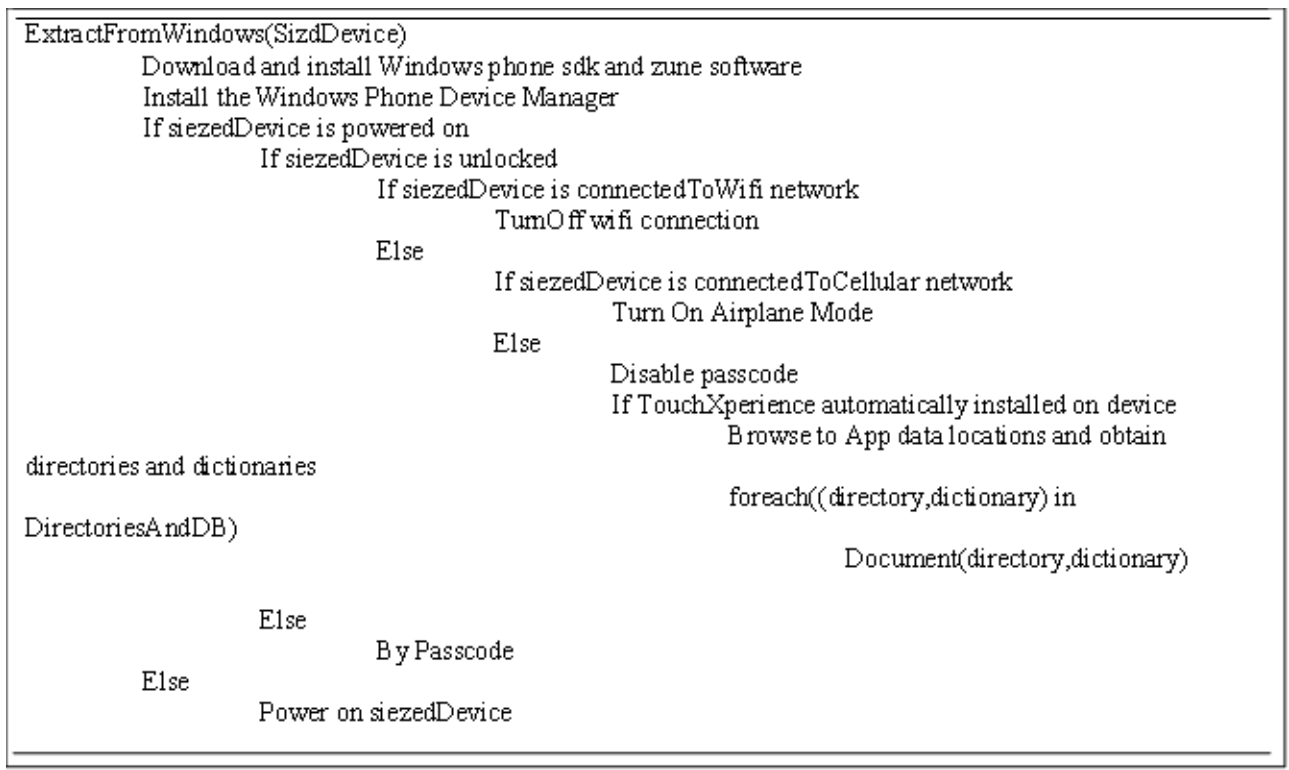

Figure 6. Extraction algorithm for Window OS.

## 2.3 Validation of the model

The developed model was validated using two approaches: (1) by experts for its applicability and functionality, these experts were purposely selected from information technology, information security and computer forensic and network security fields, law enforcement agencies, solution developers as well as researchers in the field of computer and digital forensics. The second approach was through comparison with the previous models in literature.

### 2.3.1 Applicability and Functionality of the Model

Descriptive statistics were used to determine the applicability of the model in measuring the state of digital forensic evidence extraction process models for mobile devices based on the feedback from the experts in the area of digital forensic evidence extraction process models for mobile devices. In order to determine the applicability and functionality of the model, descriptive statistics were used as shown in Table 3.

**Table 3.** Applicability in the digital forensic evidence extraction process for mobile devices

| Variables | SD/D/NS Frequency | SD/D/NS Percent (%) | A/SA Frequency | A/SA Percent (%) |
|---|---|---|---|---|
| 1. Do you understand this model with ease? | 3 | 20 | 12 | 80 |
| 2. Can you use/apply this model with ease? | 3 | 20 | 12 | 80 |
| 3. Do you consider the factors leading to measuring of digital forensic evidence extraction process model logically arranged? | 0 | 0 | 15 | 100 |
| 4. Is the explanation of the various modules within this model clear? | 4 | 26.7 | 11 | 73.3 |
| 5. Is there independence among these modules? | 2 | 13.7 | 13 | 86.3 |
| 6. Does the model guide the measuring of digital forensic evidence extraction process models for mobile devices? | 0 | 0 | 15 | 100 |
| **Average** | 0.8 | 13.4 | 13 | 86.6 |

SD= Strongly Disagree, D = Disagree, NS= Not Sure, A= Agree and SA= Strongly Agree

Analysis of all the items under the applicability of the developed model shows that almost all the 12/15 (86.6%) participants confirmed the applicability of that the developed Digital Forensic Evidence Extraction Model in guiding digital forensic evidence extraction process models for mobile devices. On the other hand, only 3 of the 15 (13.4%) had a different view about the applicability of this model in digital forensic evidence extraction process models for mobile devices. The results largely show the model applicability in the digital forensic evidence extraction process for mobile devices with 86.6% embracing it. On the other hand, the functionality of the developed Digital Forensic Evidence Extraction Process Model regarding the ease of use, (6.4%) of the respondents had a positive view about the model ease of use and that several modules within the model show interactivity (27.7%). In the same way (8.5%), confirmed independence among the several modules within the model and that the model is applicable in the digital forensic evidence extraction process for mobile devices (8.5%), and that it uses a simple language (12.8%). With (27.7%) putting their weight behind the model being used for guiding the extraction of digital evidence as depicted in Table 4.

**Table 4.** Model functionality in the digital forensic evidence extraction process for mobile devices

| Variables | Frequency | Percent |
|---|---|---|
| 1. Can you use this model with ease? | 3 | 6.4 |
| 2. Is there interactivity of the various modules within this model? | 13 | 27.7 |
| 3. Is there independence among these modules? | 4 | 8.5 |
| 4. Is the model above applicable in a developing country? | 4 | 8.5 |
| 5. Is the model easy to understand? | 5 | 10.6 |
| 6. Does it use simple language? | 6 | 12.8 |
| 7. Does the model guide measurement of digital forensic evidence extraction process models for mobile devices? | 13 | 27.7 |

### 2.3.2 Comparison analysis

A comparative analysis was done between this developed metric and model with existing models and metrics discussed in the literature and this model was found to outweigh the models discussed in literature thereby qualifying the metric and model a consistent metric that can be used for the extraction of digital forensic evidence in mobile devices run by the four operating system platforms selected for this study that is android, windows, apple iOS and Blackberry operating system as seen in Table 5.

**Table 5.** Comparative analysis of existing models with the proposed model

| Process/Phases in the Proposed model | NIST Guidelines | HDFI model | DEFSOP | SDFIM | MFP | SFIM | DFRWS |
|---|---|---|---|---|---|---|---|
| Device status check | ✓ | ✓ | ✓ | ✓ | ✓ | ☐ | ☐ |
| Preparation | ☐ | ✓ | ✓ | ✓ | ☐ | ✓ | ✓ |
| Identify evidence | ✓ | ✓ | ☐ | ✓ | ✓ | ✓ | ✓ |
| Recover data | ☐ | ☐ | ☐ | ☐ | ☐ | ☐ | ☐ |
| Forensic analysis | ✓ | ☐ | ✓ | ✓ | ✓ | ✓ | ✓ |
| Verification | ☐ | ☐ | ☐ | ☐ | ☐ | ☐ | ☐ |
| Documentation | ✓ | ✓ | ☐ | ☐ | ☐ | ✓ | ☐ |

NIST-National Institute of Science and Technology, HDFI-Harmonized Digital Forensic investigation, DEFSOP- Digital Evidence Forensic Standard Operating Procedure, SDFIM- Systematic Digital Forensic Investigation Model, MEP- Modeling the Forensic Process, SFIM- Smartphone Forensic investigation model, DFRWS-,

Based on steps or phases involved in the process models reviewed, it can be concluded the proposed model is most suitable because it summarizes most of the phases and steps proposed in earlier models and reveals the complexities in the models reviewed for example a look at the National Institute of Science and Technology (NIST) Guidelines shows that it has very limited steps; therefore, they are not appropriate enough for performing digital evidence extraction thoroughly. The Harmonized Digital Forensic investigation model presents preparation, identification, and documentation stages which this proposed model also addresses however a critical consideration of device status check are ignored in this model, forensic analysis, recovery of data, and verification which are key concerns in digital evidence extraction have also not been addressed.

Though the Digital Evidence Forensic Standard Operating Procedure, The Systematic Digital Forensic Investigation Model, and Modeling the Forensic Process all present several phases or steps to be followed, it can be noted that there are several repetitions in these stages and all of them concentrate more on the investigation itself other than extraction which the proposed model addresses right from device seizure to evidence extraction.

The Smartphone Forensic investigation model is close to the proposed model except that it concentrated more on the investigation other than evidence extraction and critically lacks the device status check and data recovery phases as pointed out in the proposed model as one of the key critical issues in digital evidence extraction in mobile devices.

## 3. Results and Discussion

### 3.1 Reliability testing

The Cronbach's α value of the various constructs ranging from 0.591 to 0.850 demonstrated the ability of the measure of internal consistency of the constructs used in this study by ensuring that none of the constructs fell below the moderate to high-reliability test. The predictive power of the regression model of this study, with adjusted R-Square 0.848 indicates an appropriate level of explained variance [28]. This implies that the independent variables and constructs used in this study are significant in understanding the causes of digital evidence extraction process model inconsistencies in mobile devices running various operating systems platforms. For example from the results of the study, it was observed that extraction methods used during evidence extraction and analysis like whether the examiner applied logical, manual, physical, or brute force approach during the process of examining a mobile device will play a significant role in ensuring the issues of consistency, similarly, forensic documentation process came out as a key contributor in ensuring consistency in the processes followed during evidence extraction whereby certain stages or phases in the extraction process ought to be documented if the results are to be repeatable and defensible in courts of law. This, therefore, justifies the choices of the constructs used in this study having support from literature and therefore, the results of this study hence, generate several issues that may be of interests to ICT practitioners, Researchers, Law enforcement authorities, Regulatory Authorities, and Business community to have a clear understanding of the factors that cause inconsistencies in digital forensics evidence extraction in mobile devices [19, 38–40]. Once these factors are clearly understood, factoring them during solution development for solution developers, paying attention to them during an investigation by forensic examiners or investigators would aid the process of collecting, preserving, and presenting evidence to courts of law for law enforcement agencies.

### 3.2 Descriptive statistics for the constructs

The descriptive statistics as shown in Table 6, give a clear view of how these constructs were ranked based on the mean responses with PF coming out significantly with a mean response of 4.36 followed by the FDP and FET had the least mean response. This implies that where there are clear policies regarding the handling, acquisition, preservation, documentation, and presentation of digital pieces of evidence, then there should be minimal inconsistencies in the mobile devices' digital evidence extraction process model. This is followed by the forensics documentation process suggesting concurrence with recent studies citing a lack of clear technical documentation of existing process models of mobile devices digital evidence extraction methods [6]. Forensic extraction tools are ranked last among the eight constructs, this could be attributed to the fact that there are several digital evidence extraction tools and most investigators face challenges in choosing the correct tool for digital evidence extraction in mobile devices depending on the mobile device platform they are encountered with [20].

**Table 6.** Descriptive Statistics for constructs and their rankings

| Construct | N | Mean | Std. Deviation | Rank |
|---|---|---|---|---|
| Policy Factors (PF) | 85 | 4.36 | .386 | 1 |
| Device Factors (DF) | 85 | 4.21 | .556 | 2 |
| Forensic Documentation Process (FDP) | 85 | 4.11 | .434 | 3 |
| Data Type Factors (DTF) | 85 | 4.11 | .564 | 4 |
| Extraction Method Factors (EM) | 85 | 4.01 | .456 | 5 |
| Nature of Data (ND) | 85 | 3.90 | .624 | 6 |
| operating System Platform (MDF) | 85 | 3.80 | .855 | 7 |
| Forensic Extraction Tools (FET) | 85 | 3.08 | .946 | 8 |
| Valid N (listwise) | 85 | | | |

### 3.2.1 Policy factor (PF)

The means and standard deviations of aggregated measures for the seven items used to measure the PF construct are illustrated in Table 7. In this table, seven items are used to measure this construct ranging from PF1- PF7.

**Table 7.** Descriptive statistics for Policy construct.

| Item | Mean | Std. Deviation | N |
|---|---|---|---|
| PF1 | 4.64 | .574 | 85 |
| PF2 | 4.31 | .637 | 85 |
| PF3 | 4.40 | .876 | 85 |
| PF4 | 4.32 | .582 | 85 |
| PF5 | 4.49 | .684 | 85 |
| PF6 | 4.09 | .750 | 85 |
| PF7 | 4.31 | .887 | 85 |

A strong agreement was made for the policy factor construct with the average score of (Mean = 4.36, Standard Div = 4.99) with the item on establishment of policy guidelines being the most agreed upon, PF1 (M = 4.64, SD = 0.574), followed by Training of staff on current mobile devices digital forensic evidence technologies has a positive effective on digital forensics evidence extraction inconsistencies (PF5) (M = 4.49, Standard Div = .684), Establishment of mobile devices digital forensic evidence handling unit within the organization reduces inconsistencies in mobile devices digital forensics evidence extraction (PF3) (M = 4.40, Standard Div = .876), Recruitment of qualified staff to handle mobile devices digital forensic evidence has a positive effect on evidence extraction inconsistencies PF4 (M = 4.32, Standard Div = .582), Effective implementation of policies leads to a consistent digital forensic evidence extraction process PF2 (M = 4.31, Standard Div = .637), and PF6 (M = 4.09, Standard Div = .750) being the least agreed upon item for this construct. The average inter-item correlation determines the reliability of the construct, hence, the higher the average inter-item correlation, the higher the construct reliability coefficient, Cronbach's alpha (α), subject to holding the number of items constant [28]. Table 8 shows the inter-item correlation for the items used to measure the policy factor (PF) construct.

**Table 8.** Inter- item correlation matrix for Policy Factors (PF) constructs

| Item | PF1 | PF2 | PF3 | PF4 | PF5 | PF6 | PF7 |
|---|---|---|---|---|---|---|---|
| PF1 | 1.000 | .211 | .554 | .173 | .161 | .081 | .081 |
| PF2 | .211 | 1.000 | .141 | .217 | .059 | .064 | .170 |
| PF3 | .554 | .141 | 1.000 | .168 | .203 | .232 | .132 |
| PF4 | .173 | .217 | .168 | 1.000 | .050 | .395 | .179 |
| PF5 | .161 | .059 | .203 | .050 | 1.000 | .024 | .062 |
| PF6 | .081 | .064 | .232 | .395 | .024 | 1.000 | .243 |
| PF7 | .081 | .170 | .132 | .179 | .062 | .243 | 1.000 |

Most of the items had acceptable inter-item correlation ($r>=0.2$). The least agreed upon items, namely Enactment of laws governing mobile devices digital forensic evidence extraction has a positive effect on evidence extraction inconsistencies (PF6) and Developing strategies and frameworks for investigation of mobile devices digital forensic evidences has a positive effect on evidence extraction inconsistencies in mobile devices (PF7) were also the least correlated with the rest of the items, while Establishment of policy guidelines for mobile devices digital forensic evidence extraction leads to a consistent digital forensics evidence extraction process PF1, Establishment of mobile devices digital forensic evidence handling unit within the organization reduces inconsistencies in mobile devices digital forensics evidence extraction PF3 and Recruitment of qualified staff to handle mobile devices digital forensic evidence has a positive effect on evidence extraction inconsistencies PF4 were positively correlated with the rest of the items for the construct. There was a moderate relationship ($r>=0.55$) between the establishment of policy guidelines for mobile devices digital forensic evidence extraction leads to a consistent digital forensics evidence extraction process (PF1) item, and establishment of mobile devices digital forensic evidence handling unit within the organization reduces inconsistencies in mobile devices digital forensics evidence extraction (PF3) ($r=0.55$), as well as a low correlation between recruitment of qualified staff to handle mobile devices digital forensic evidence has a positive effect on evidence extraction inconsistencies (PF4) and Enactment of laws governing mobile devices digital forensic evidence extraction has a positive effect on evidence extraction inconsistencies (PF6) ($r=0.395$). We can accordingly, conclude that the items selected for measuring the policy factor (PF) were appropriate for the measure.

### 3.2.2 Device factor

The means and standard deviations of aggregated measures for the three items used to measure the Device Factor (DF) construct are presented in Table 9.

**Table 9.** Descriptive statistics for Device factors construct items

| Item | Mean | Std. Deviation | N |
|---|---|---|---|
| DF1 | 4.45 | .627 | 85 |
| DF2 | 4.27 | .662 | 85 |
| DF3 | 4.09 | .908 | 85 |
| DF4 | 4.04 | .957 | 85 |

The strong agreement was made for the Device factor construct (DF) with the average score of (Mean = 4.21, Standard Div = 3.15) with the item on status of the mobile device during evidence intake being the most agreed upon, The status of the mobile device during evidence intake DF1 (Mean = 4.45, Standard Div = .627), followed by The type of mobile devices DF2 (Mean = 4.27, Standard Div = .662), The release versions of mobile devices DF3 (Mean = 4.09, Standard Div = .908), and DF4(Mean = 4.04, Standard Div = .957) being the least agreed upon an item for this construct. Table 10 shows the inter-item correlation for the items used to measure the Device

Factor (DF) construct. As observed, most of the items had acceptable inter-item correlation (r>=0.2). The least agreed-upon item was the type of mobile device (DF2) and was the least correlated with the release versions of mobile devices (DF3) (r=0.155). There was a moderate relationship (r>=0.568) between the type of mobile devices (DF2) item, and device connection settings (DF4) (r=0.331), as well as a low correlation between the status of the mobile device during evidence intake (DF1) and (DF2), and (DF3) with (r >0.279 but < 0.386). We can accordingly, conclude that the items selected for measuring the Device Factor (DF) were appropriate for the measure of this construct [28, 41].

Table 10. Inter-item correlation for the Device Factor (DF) construct

| Item | DF1 | DF2 | DF3 | DF4 |
|---|---|---|---|---|
| DF1 | 1.000 | .279 | .385 | .331 |
| DF2 | .279 | 1.000 | .155 | .568 |
| DF3 | .385 | .155 | 1.000 | .229 |
| DF4 | .331 | .568 | .229 | 1.000 |

### 3.2.3 Extraction method factor

The means and standard deviations of aggregated measures for the ten items used to measure the Extraction Method Factors (EMF) construct. From Table 11, a strong agreement was made for the Extraction Method factor construct with the average score of (Mean = 4.12, Standard Div = 0.83) with the item on Physical acquisition being the most agreed upon, EMF3 (Mean = 4.46, Standard Div = 0.716), followed by EMF1 (Mean = 4.39, SD = 0.773), EMF5 (Mean = 4.14, Standard Div = 0.789), Logical acquisition EMF2 where (Mean = 4.11, Standard Div = 0.772), Brute force acquisition EMF4 (Mean = 3.96, SD = 0.763), and Architecture EMF6 (Mean = 3.91, Standard Div = 0 .959) being the least agreed upon item for this construct.

Table 11. Descriptive statistics for Extraction Method factors construct items

| Item | Mean | Std. Deviation | N |
|---|---|---|---|
| EMF1 | 4.39 | .773 | 85 |
| EMF2 | 4.11 | .772 | 85 |
| EMF3 | 4.46 | .716 | 85 |
| EMF4 | 3.96 | .763 | 85 |
| EMF5 | 4.14 | .789 | 85 |
| EMF6 | 3.91 | .959 | 85 |
| EMF7 | 3.93 | 1.021 | 85 |
| EMF8 | 4.25 | 1.022 | 85 |
| EMF9 | 4.09 | .840 | 85 |
| EMF10 | 3.85 | 1.160 | 85 |

Similarly, in Table 12 the Inter-item correlation for the different factors and most of the items had acceptable inter-item correlation (r>=0.2). The least agreed-upon item was Architecture (EMF6), File system (EMF8), Data storage mechanism (EMF9), and Instant messaging applications (EMF10). Subsequently, they were the least correlated with Manual acquisition (EMF1), Logical acquisition (EMF2), Physical acquisition (EMF3) with (r<=0.2). There was a moderate relationship (r>=0.589) between (EMF1) and EMF2, as well as a low correlation between (EMF3) and (EMF2), and (EMF10) with (r >0.2 but < 0.386). We can accordingly, conclude that the items selected for measuring the Extraction Method factors (EMF) were appropriate for the measure this construct.

Table 12. Inter-item correlation for Extraction Method Factors (EMF) construct

| Item | EMF1 | EMF2 | EMF3 | EMF4 | EMF5 | EMF6 | EMF7 | EMF8 | EMF9 | EMF10 |
|---|---|---|---|---|---|---|---|---|---|---|
| EMF1 | 1.000 | .589 | .126 | .266 | .143 | -.014 | .322 | -.002 | .016 | .001 |
| EMF2 | .589 | 1.000 | .213 | .593 | .190 | -.002 | .251 | .087 | .021 | .071 |
| EMF3 | .126 | .213 | 1.000 | .357 | .411 | -.058 | -.183 | -.092 | .086 | .100 |
| EMF4 | .266 | .593 | .357 | 1.000 | .345 | .174 | .088 | .057 | .135 | .236 |
| EMF5 | .143 | .190 | .411 | .345 | 1.000 | .159 | .367 | .325 | .303 | .050 |
| EMF6 | -.014 | -.002 | -.058 | .174 | .159 | 1.000 | .370 | .230 | .159 | .030 |
| EMF7 | .322 | .251 | -.183 | .088 | .367 | .370 | 1.000 | .348 | .063 | -.200 |
| EMF8 | -.002 | .087 | -.092 | .057 | .325 | .230 | .348 | 1.000 | .111 | .153 |
| EMF9 | .016 | .021 | .086 | .135 | .303 | .159 | .063 | .111 | 1.000 | .394 |
| EMF10 | .001 | .071 | .100 | .236 | .050 | .030 | -.200 | .153 | .394 | 1.000 |

### 3.2.4 Nature of Data factors

The means and standard deviations of aggregated measures for the five items used to measure the Nature of Data Factors (ND) construct are illustrated in Table 13.

Table 13. Descriptive statistics for Nature of Data factors

| Item | Mean | Std. Deviation | N |
|---|---|---|---|
| ND1 | 4.32 | .790 | 85 |
| ND2 | 3.79 | .773 | 85 |
| ND3 | 4.19 | .794 | 85 |
| ND4 | 3.69 | .859 | 85 |
| ND5 | 3.53 | 1.042 | 85 |

A strong agreement was made for the Nature of Data factor construct with the average score of (Mean = 3.90, Standard Div = 0.851) with the item on internal and visible data being the most agreed upon, ND1 (Mean = 4.32, Standard Div = 0.790), followed by External and visible ND3 (Mean = 4.19, SD = 0.794), Internal but hidden ND2 (Mean = 3.79, Standard Div = 0.773), External but hidden ND4 (Mean = 3.69, Standard Div = 0.859), and Encrypted Data ND5 (Mean = 3.53, Standard Div = 1.04) being the least agreed upon item for this construct.

Table 14. Inter-item correlation for Nature of Data Factors (ND)

| Item | ND1 | ND2 | ND3 | ND4 | ND5 |
|---|---|---|---|---|---|
| ND1 | 1.000 | .540 | .549 | .355 | .328 |
| ND2 | .540 | 1.000 | .395 | .421 | .303 |
| ND3 | .549 | .395 | 1.000 | .295 | .353 |
| ND4 | .355 | .421 | .295 | 1.000 | .648 |
| ND5 | .328 | .303 | .353 | .648 | 1.000 |

Regarding Table 14, Most of the items had acceptable inter-item correlation (r>=0.2). The least agreed-upon item was Encrypted Data ND5 and was the least correlated with External but hidden ND4 (r =0.353). There was a moderate relationship (r>=0.540) between Internal and visible (ND1) item, and External and visible ND3 (r>=0.549), as well as a low correlation between External but hidden (ND4) and External and visible (ND3) with (r <=0.295). We can accordingly, conclude that the items selected for measuring the Nature of Data Factor (ND) were appropriate for the measure.

### 3.2.5 Correlation of individual OS and Constructs

Table 15 shows how different individual operating system platforms relate with various constructs; from this table it indicates that the Forensic Documentation process (FDP) has a significant correlation with Apple iOS, at .404 (40.4%), closely followed by Android at .268 (26.8%), windows at .229 (22.9%) while Blackberry had the least significant correlation at .008 (0.08%). Forensics Extraction tools (FET) posted significant correlation with Apple iOS and Blackberry operating system at .524(52.4%) and .667(66.7%) respectively, Windows came third with .285 (28.5%) and Android trailed with .178 (17.8%). Data type showed the least correlation across all the four operating system platforms closely followed by Policy factors. The implication is that FET, FDP, EM and ND are more significant factors in understanding how they influence evidence extraction in mobile devices running such operating system platform, while given the fact that any of the four operating systems can have either the same or different kinds of data for example call logs, browser history, short message services or videos could explain the reason why data type posted the least significant correlation.

**Table 15.** Correlation of individual operating systems and independent constructs

| Item | | P F | D F | E M | DTF | N D | FET | FDP |
|---|---|---|---|---|---|---|---|---|
| Android | Pearson Correlation | -.034 | .101 | .210 | .036 | .130 | .178 | .268* |
| | Sig. (2-tailed) | .755 | .357 | .054 | .741 | .237 | .104 | .013 |
| | N | 85 | 85 | 85 | 85 | 85 | 85 | 85 |
| Window | Pearson Correlation | .132 | .199 | .421** | .221* | .236* | .285** | .229* |
| | Sig. (2-tailed) | .230 | .068 | .000 | .042 | .030 | .008 | .035 |
| | N | 85 | 85 | 85 | 85 | 85 | 85 | 85 |
| Apple iOs | Pearson Correlation | .073 | .364** | .496** | .032 | .318** | .524** | .404** |
| | Sig. (2-tailed) | .507 | .001 | .000 | .768 | .003 | .000 | .000 |
| | N | 85 | 85 | 85 | 85 | 85 | 85 | 85 |
| Blackberry operating system | Pearson Correlation | -.116 | .190 | .306** | -.221* | .325** | .667** | .008 |
| | Sig. (2-tailed) | .291 | .081 | .004 | .042 | .002 | .000 | .944 |
| | N | 85 | 85 | 85 | 85 | 85 | 85 | 85 |

PF-Policy Factor, DF- Device Factors, EM-Extraction Method, DTF-Data Type Factors, ND- Nature of Data, FET-Forensic Extraction Tool and FDP-Forensic Documentation Process

### 3.3 Regression analysis

According to Perry et al. [28], "linear regression (LR) is a method used to model the linear relationship between a dependent variable and one or more independent variables. The dependent variable is sometimes also called the predict and, and the independent variables as the predictors. Linear regression is based on least-squares: the model is fit such that the sum-of-squares of differences of observed and predicted values is minimized based on six basic assumptions. The regression analysis was performed with consistency metric (CM) as the dependent variable and constructs such as (Extraction methods (EM), Forensic Extraction Tools (FET), Policy Factors (PF), Device Factors (DF), Nature of Data (ND) and Data type factors (DTF)) as independent variables. From the analysis, a significant model emerged with adjusted R-Square .848 which represents 84.8% and hence the Predictor variable included in the analysis was found to be significant. The overall F =79.238, Sig. .000b about extraction methods, data type, nature of data, policy factors, forensic documentation, forensic extraction tools, device factors. The results signify that the model is statistically significant, valid, and fit. The validity of the model signifies that the consistency metric predicts a significant relationship with extraction inconsistencies. Henceforth, the model was fit enough to be based on making conclusions and recommendations.

The regression model reveals Adjusted R Square =0.848 which signifies the consistency metric is greatly influenced by these factors such as policy, data type, nature of data, extraction method, and forensic documentation process. With an adjusted R square of 0.848 which represents 84.8% of the constructs used in predicting the consistency metric to be used in the evidence extraction process model from those mobile devices running on the four mobile operating systems, that is android, windows, apple iOs, and blackberry operating system. This model fit confirms what literature revealed about factors such as documentation, extraction methods being majoring causes of inconsistencies in mobile devices evidence extraction process models [4, 36–38], subsequently, other factors such as policy [32, 39, 40], nature of data [41] and data type [15, 42] have a minor contribution to evidence extraction process inconsistencies. From this Table 16, two factors have come out very significant that is extraction method factors standing at B = 1.030 and Device factors at B= .078, these positive values indicate that as independent variables increase consistency metric as a dependent variable also increase this is supported by the literature. The coefficient of the regression analysis also indicates that as some of the independent variables increases, consistency decreases, and standard error decrease for example nature of data B = -0.029 and Beta = -0.037 with sig. at 0.443. The implication here is that these factors do not significantly contribute to the consistency metric and therefore they have less influence on the consistency process model during evidence extraction in mobile devices running the four operating system platforms used in this study.

**Table 16.** Coefficient of regression analysis with consistency metric as the dependent variable

| Model | | Unstandardized Coefficients B | Std. Error | Standardized Coefficients Beta | T | Sig. |
|---|---|---|---|---|---|---|
| 1 | (Constant) | -.649 | .347 | | -1.870 | .065 |
| | EM | 1.030 | .054 | .897 | 18.963 | .000 |
| | ND | -.029 | .037 | -.037 | -.771 | .443 |
| | DTF | .009 | .039 | .011 | .231 | .818 |
| | EMF | .016 | .046 | .016 | .346 | .730 |
| | DF | .078 | .039 | .092 | 1.975 | .052 |
| | PF | .040 | .050 | .035 | .797 | .428 |

The findings of this study showed that Forensics Extraction tools, Extraction methods, Nature of Data, Device type, and Forensics Documentation Process are the primary contributors to Extraction inconsistencies. These findings support results from recent studies that revealed that discrepancies in recovering and reporting the data residing on a device have been noted in the previous testing of tools and updates or new versions of a tool. This is consistent with interview findings that revealed that the type of data, nature of data, extraction method, are among the major contributors of inconsistencies in mobile devices forensic evidence extraction models. Further, the study findings established that the Policy factor is a metric for the specification of a consistent digital forensic evidence extraction model for mobile devices based on Android, Windows, Apple iOS, and Blackberry OS. Moreover, the Device factor is part of metrics for the specification of a consistent digital forensic evidence extraction model for mobile devices based on Android, Windows, Apple iOS, and Blackberry Operating System (OS).

The present study shown that the Extraction Method factor is a metric for the specification of a consistent digital forensic evidence extraction model for mobile devices based on Android, Windows, Apple iOS, and Blackberry OS. The study findings established that the Nature of Data factors is metrics for the specification of a consistent digital forensic evidence extraction model for mobile devices based on Android, Windows, Apple iOS, and Blackberry Operating System. This is in handy with Brian Cusack [43] who posits that the high-level process of digital forensics includes the acquisition of data from a source, analysis of the data and extraction of evidence, and preservation and presentation of the evidence. From this study, it was noticed that Forensic Extraction tools are metrics for the specification of a consistent digital forensic evidence extraction model for mobile devices based on Android, Windows, Apple iOS, and Blackberry operating systems. While Forensic Documentation process forms part of metrics for the specification of a consistent digital forensic evidence extraction model for mobile devices.

**Conclusion**

The extraction process models developed borrowed the principles of consistencies, repeatability, and standardization as presented in earlier studies of the generalized forensic framework from previous studies, this model goes further to enumerate sequentially each step that should be followed in evidence extraction for each of the mobile operating systems thereby ensuring that there are consistencies at every step of the extraction process. These sequential or chronological steps (stages) if followed will yield positive results across the four mobile operating systems and it is believed that this model can act as a standard for any other mobile operating system platform that has not been part of this study, considering that the architecture of mobile devices does not differ significantly in terms of storage, processing, and application. The Smartphone Forensic investigation model is close to the proposed model except that it concentrated more on the investigation other than evidence extraction and critically lacks the device status check and data recovery phases as pointed out in the proposed model as one of the key critical issues in digital evidence extraction in mobile devices. Future work should focus on practically testing these models and comparing the results for consistencies across different operating system platforms